\documentclass[letterpaper]{article}
\usepackage{kmn}
\usepackage{epsfig,subfigure}
\title[8--13$\mu$m Dust Emission Features in Galactic Bulge Planetary
Nebulae]{8--13$\mu$m Dust Emission Features in Galactic Bulge
Planetary Nebulae}
\author[S.~Casassus et al.]
  {S.~Casassus$^{1,2}$, P.\,F.~Roche$^1$, D.\,K.~Aitken$^3$ \& C.\,H.~Smith$^4$ \\
	$^1$ Astrophysics, Physics Department, Oxford University,
  Keble Road, Oxford OX1 3RH\\
 	$^2$ Departamento de Astronom\'{\i}a, Universidad de Chile,
  Casilla 36-D, Santiago, Chile.\\
	$^3$ Department of Physical Sciences, University of Hertfordshire, Hatfield, Herts AL10 9AB\\
	$^4$ School of Physics, University College, UNSW, Canberra,
  ACT 2600, Australia.}

\date{Accepted ... Received ...}
\pagerange{\pageref{firstpage}--\pageref{lastpage}}
\pubyear{2000}
\def\gs{\mathrel{\raise1.16pt\hbox{$>$}\kern-7.0pt
\lower3.06pt\hbox{{$\scriptstyle \sim$}}}}
\def\ls{\mathrel{\raise1.16pt\hbox{$<$}\kern-7.0pt
\lower3.06pt\hbox{{$\scriptstyle \sim$}}}}
\begin{document}
\label{firstpage}
\maketitle
\begin{abstract}
A sample of 25 IR-bright planetary nebulae (PNe) towards the Galactic
bulge is analysed through 8--13$\mu$m spectroscopy. The classification
of the warm-dust emission features provides a measure of the C/O
chemical balance, and represents the first C/O estimates for bulge
PNe.  Out of 13 PNe with identified dust types, 4 PNe have emission
features associated with C-based grains, while the remaining 9 have
O-rich dust signatures.  The low fraction of C-rich PNe, $\ls$30\%,
contrasts with that for local PNe, around $\sim$80\%, although it
follows the trend for a decreasing frequency of C-rich PNe with
galactocentric radius (paper~I). We investigate whether the PNe
discussed here are linked to the bulge stellar population (similar to
type~IV, or halo, PNe) or the inner Galactic disk (a young and
super-metal-rich population). Although 60\% of the PNe with warm dust
are convincing bulge members, none of the C-rich PNe satisfy our
criteria, and they are probably linked to the inner Galactic disk.  In
the framework of single star evolution, the available information on
bulge PNe points towards a progenitor population similar in age to
that of local PNe (type I PNe are found in similar proportions), but
super-metal-rich (to account for the scarcity of C-rich objects). Yet
the metallicities of bulge PNe, as inferred from [O/H], fail to reach
the required values - except for the C-rich objects. It is likely that
the sample discussed here is derived from a mixed disk/bulge
progenitor population and dominated by type IV PNe, as suggested by
Peimbert (1992).  The much higher fraction of O-rich PNe in this
sample than in the solar neighbourhood should result in a
proportionally greater injection of silicate grains into the inner
Galactic medium.  
\end{abstract}

\section{Introduction.}

Planetary nebula (PN) compositions reflect the initial composition and
nuclear processing undergone by their progenitors, and provide
valuable information on the end point of evolution for populations of
intermediate mass stars in various galactic environments. Ratag et
al. (1997) have published an extensive abundance analysis of PNe
towards the Galactic bulge, providing helium, oxygen and nitrogen
abundances.  The results are surprising as many bulge PNe have high
N/O ratios, which is characteristic of Peimbert type I PNe.  The
temptation to link the high N/O ratio in the bulge with the high
progenitor masses of type I PNe, in excess of 2--3\,M$_{\odot}$, is
countered by growing evidence that the bulge and halo are coeval
(Ortolani 1995).  In an addendum to his review on PNe, Peimbert (1992)
favours a link with halo PNe (type IV, defined by a very low
metallicity), which sometimes have N/O ratios within the range of type
I PNe. However, carbon is conspicuously absent from the previous
abundance analyses, and the C/O chemical balance is also a function of
progenitor mass.

The C/O abundance ratio in PNe is usually calculated from gas phase
abundances, through fine structure emission lines of C and O ionised
up to three times. But the bright C lines are \hbox{[C\,{\sc
ii}]}~$\lambda2326$, \hbox{[C\,{\sc iii}]}~$\lambda1908$,
\hbox{[C\,{\sc iv}]}~$\lambda1550$, and extinction is severe at such
short wavelengths. As the lines of sight to the bulge are usually
affected by strong extinction, no C abundances are available for bulge
PNe. However, the C/O chemical balance can be inferred from the
8--13$\mu$m warm dust signatures, through the identification of the
grain type. Silicate emission corresponds to oxygen-rich environments,
SiC emission is typical of carbon-rich environments, while the
unidentified infrared (UIR) emission bands are associated with a
strong overabundance of carbon relative to oxygen (e.g. Barlow 1983,
Roche 1989, see also Casassus et al. 2001a, paper~I).

In this article we analyse a sample of bulge PNe through their dust
signatures.  We present 8--13$\mu$m spectra for a sample of 18 PNe
observed towards the Galactic bulge (which meet the bulge-membership
criteria of Acker et al.  1991), and analyse these, together with 7
objects previously reported by Roche (1987). We determine the dominant
dust emission features in terms of grain emissivities, thus deriving
the C/O chemical balances. Section \ref{sec:bulge_obs} contains a
description of the observations.  A compilation of the data available
for the sample of bulge PNe with 8--13$\mu$m spectra, together with a
discussion on the criteria for bulge membership, is presented in
Section~\ref{sec:data}.  In Section~\ref{sec:bulge_mod} we interpret
the available information on bulge PNe.  In the framework of single
star evolution, the rarity of C-rich PNe and the high frequency of
N-enrichment links the `bulge' PNe with the inner Galactic disk rather
than with the bulge stellar population.  But the O abundances in bulge
PNe do not reach the required super-metal-rich values. The failure of
single star evolution models in accounting for bulge PNe is thus
highlighted.  Section~\ref{sec:conc_bulge} summarises our conclusions.

\section{8--13$\mu\lowercase{m}$ spectra of PNE towards the Galactic
bulge}\label{sec:bulge_obs}

The objects selected were required to satisfy the following criteria,
\begin{itemize}
\item good detection by {\em IRAS} at 25~$\mu$m, and 12~$\mu$m flux in excess
of $\gs 0.5$~Jy, or high upper limits,
\item diameters less than $10''$, and almost all $<5''$,
\item listed as likely bulge PNe in the Strasbourg-ESO catalogue
(Acker et al. 1992), i.e.
$-10<l<10$, $-10<b<10$, $F(6\,\mathrm{cm})<100$~mJy.
\end{itemize}
The spectra were acquired during July 23--26 1990, using UCLS at
UKIRT, with the 40 l/mm grating and an aperture of 4.5 arcsec in
diameter. The linear array of 25 Si:As photoconductive detectors was
oversampled two times. Flux calibration was relative to standard stars
and is accurate to 20\%, and Table~\ref{table:obslog_bulge} contains a
list of the observed PNe, as well as the emission line fluxes.
\begin{table}
\caption{List of objects observed with UCLS at UKIRT in July 1990.  
Emission line fluxes are 
10$^{-15}$~W\,m$^{-2}$}\label{table:obslog_bulge}
\begin{center}
\small
\begin{tabular}{lrccc}
       &    PN\,G      & [Ar\,{\sc iii}] & [S\,{\sc iv}] & [Ne\,{\sc ii}]  \\
       &         & 8.99\,$\mu$m   & 10.52\,$\mu$m & 12.81\,$\mu$m   \\
H1-54  &    2.1--4.2  & --- &  ---	& 1.7  \\
H1-63  &    2.2--6.3  & --- &  ---	& 0.5  \\
Cn1-5  &    2.2--9.4  & 0.7 & 2.0	& 1.2 \\
M1-38  &    2.4--3.7  & --- & ---	& 3.9\\
M1-35  &    3.9--2.3  & 1.8 & 8.7	& 1.0 \\
M2-29  &    4.0--3.0  & --- & 0.4  & --- \\
H1-53  &    4.3--2.6  & 1.4 & 1.5  & 1.7 \\
M1-25  &    4.9+4.9  & 2.0 & ---  & 8.6 \\
M1-20  &    6.1+8.3  & 0.7 & 1.2  &  --- \\
M3-15  &    6.8+4.1  & 1.3 & 4.9 & --- \\
M3-21  &  355.1--6.9  & 0.8 & 7.3 & --- \\
H1-32  &  355.6--2.7  & --- & 0.9 & --- \\
M1-27  &  356.5--2.3  & --- & --- & 5.8 \\
M2-24  &  356.9--5.8  & --- & --- & --- \\
M3-38  &  356.9+4.4  & --- & 2.2 & 0.3 \\
M3-8   &  358.2+4.2  & 0.6 & 1.0 & 0.9 \\
Th3-26 &  358.8+3.0  & --- & 3.3 & --- \\
M1-29  &  359.1--1.7  & 1.9 & 3 & 0.7 
\end{tabular}
\end{center}
\end{table}

\begin{figure}
\resizebox{8.5cm}{!}{\epsfig{file=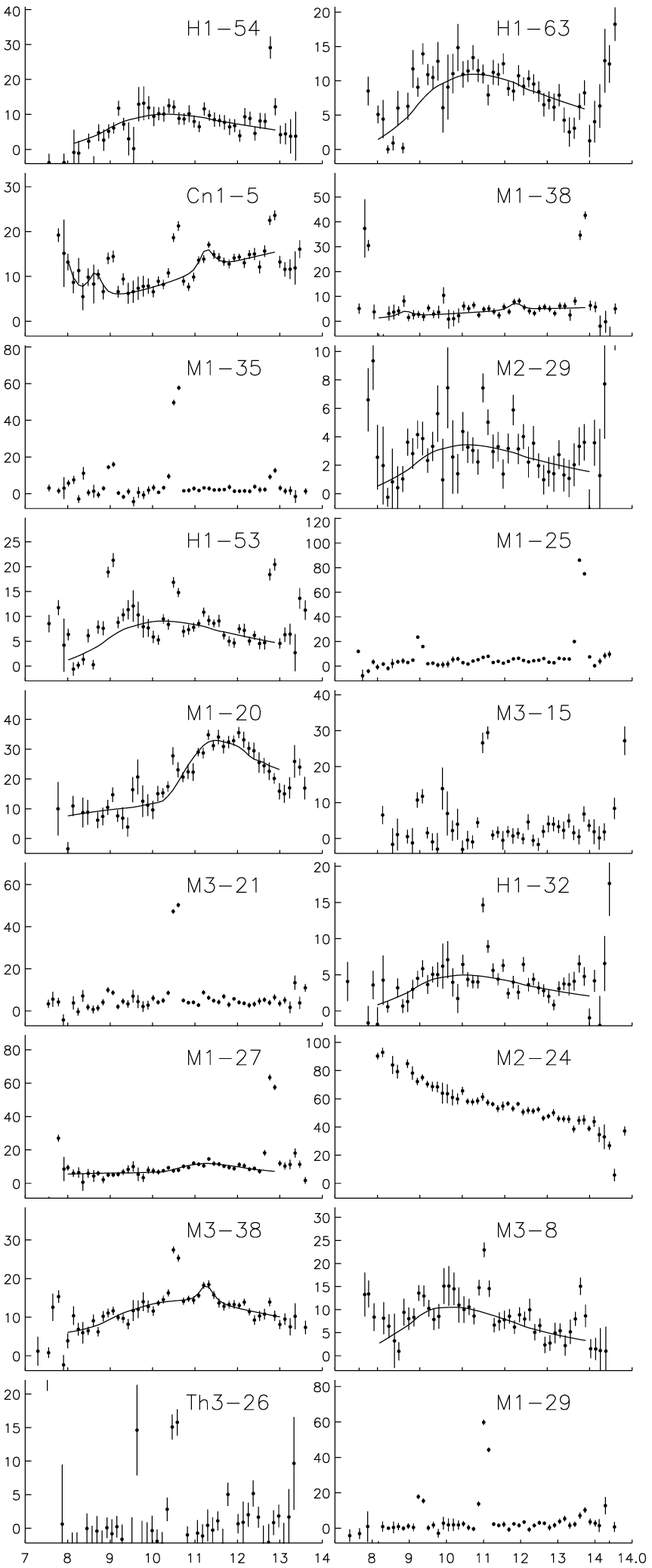}}
\caption{10$\mu$m spectra of the PNe listed in Table
\ref{table:obslog_bulge}. The abscissae are the wavelengths in
$\mu$m, and the ordinates the flux density in
10$^{-15}$W\,m$^{-2}$\,$\mu$m$^{-1}$. Fits to the dust emission 
continua are shown by the solid lines (fit parameters are in
Table~\ref{table:dusttypes_bulge}). Bright emission lines of [A\,{\sc iii}],
[S\,{\sc iv}] or [Ne\,{\sc ii}] at 9.0, 10.5 and 12.8~$\mu$m are excluded from the
fitting procedure.}\label{fig:dustspecs_bulge} 
\end{figure}

Fig.~\ref{fig:dustspecs_bulge} shows the resulting spectra, together
with the fits based on the different grain emissivities, according to
the procedure from Aitken et al. (1979), and also described in
paper\,I. Table~\ref{table:dusttypes_bulge} lists the warm dust
identification for the 25 bulge PNe which have been observed
spectroscopically at 8--13$\mu$m (henceforth the bulge PN sample).
The spectra for PNe in Table~\ref{table:dusttypes_bulge} which are not
listed in Table~\ref{table:obslog_bulge} are published in Roche
(1987). The column under `comments' contains the best fit parameters
to the 8--13$\mu$m continua, with the notation of paper~I, leading to
a classification of dust signatures in terms of the dominant component
(see Table~3 in paper\,I): 
\begin{equation}
F_{\lambda}={\sum}_{i} a_{i} \epsilon_i({\lambda})
B({\lambda},T_{i})/B(10{\mu}m,T_i), 
\end{equation}
where we sum over grain types, each with an emissivity $\epsilon_i$,
and where $B({\lambda},T_{i})$ is a Planck function at temperature
$T_i$. The free parameters are $a_i$ and $T_i$, the relative
contribution of each grain type being
$a^{\prime}_{i}=a_{i}/\sum_{j}a_{j}$. In this classification, a
superposition of silicate emission and the UIR bands is classified as
O-rich, because all `O' PNe classified in this way have gas phase
C/O$<1$ (in the sample discussed here, only M\,3-38 shows such a
superposition).

\begin{table*} 
 \vbox to220mm{\vfil Landscape Table to go here
 \caption{}  \label{table:dusttypes_bulge}
 \vfil}
\end{table*}

The proportion of PNe with C-based grains is smaller for PNe with
longitudes towards the bulge than in the disk population:
\begin{center}
\begin{tabular}{ccc}\hline
Bulge & \multicolumn{2}{c}{Disk} \\
	&	$R<R_\circ^{~a}$	& $R>R_\circ$\\
$31\pm13$\% &  $70\pm9$\%	& $86\pm7$\% \\  \hline
\end{tabular}

\medskip
$^{a}$ {\small $R_\circ=8.5$~kpc corresponds to the solar circle (Kerr \& Lynden-Bell 1986)}. 
\end{center}
There is a striking difference between the bulge PN sample compared
with local PNe (see paper~I for a discussion on the dust signatures in
local PNe). This may be interpreted either as support for linking the
PNe in Table~\ref{table:dusttypes_bulge} with the bulge stellar
population, or as a strong metallicity dependence in the PN
compositions. By contrast central star properties (Tylenda et
al. 1991), or other nebular abundances (Section~\ref{sec:data} below),
have not convincingly been demonstrated to reflect the broadly varying
bulge and local galactic environments.

Does the IR-bright selection criterion preferentially select silicate
nebulae? As shown in paper\,I, the fraction of flux emitted in the
12~$\mu$m band, $F(12\mu\mathrm{m})/F_{IRAS}$, is about 25\% and
independent of dust emission type. No particular bias towards silicate
nebulae is expected, and moreover the bulge sample is subject to 
the same selection effects as the disk sample in paper~I.  
In spite of a lower $F(12\mu\mathrm{m})/F_{IRAS}$
of $\sim10$\%, `weak' nebulae are found in a proportion of 48\% in the
bulge sample, against 27\% in the disk. The higher proportion of
`weak' nebulae is probably because the Galactic bulge sample is at a 
greater average distance from the sun, and hence fainter than the disk
sample, but  also many bulge PNe have upper limits only in the 12~$\mu$m
band.  

\begin{figure}
\begin{center}
\resizebox{8cm}{!}{\epsfig{file=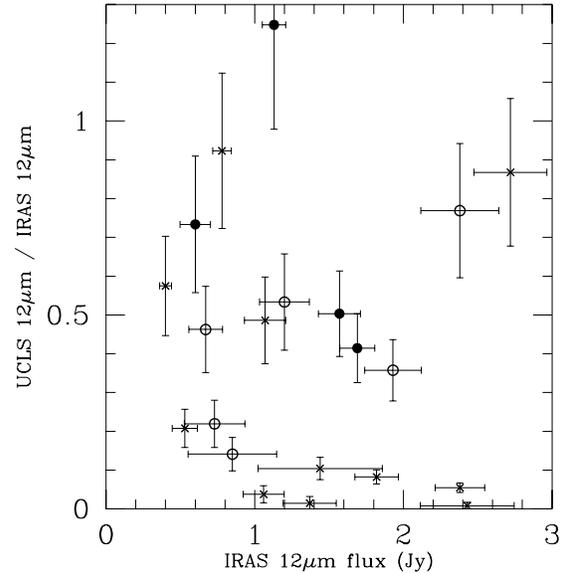}}  
\caption{The ratio of UCLS to {\em IRAS} 12~$\mu$m flux as a function
of {\it IRAS} flux. Open and filled circles correspond to  O-rich
and C-rich objects respectively, while crosses denote weak continuum
PNe.} 
\label{fig:UCLS_IRAS}
\end{center}
\end{figure}

It is noteworthy that the 11-13 $\mu$m flux measured from the ground
with the UCLS is in many cases substantially lower than that measured
by {\em IRAS}.  The average 12 $\mu$m flux ratio is 0.36, but this
varies from 0.67 for C-rich objects, 0.48 for O-rich objects and 0.17
for PNe with weak continua, as shown in Fig.~\ref{fig:UCLS_IRAS}
(objects with {\em IRAS} upper limits at 12 $\mu$m are excluded, but
those with uncertain detections are included).  While source extension beyond 
the UCLS aperture and pointing errors may account for some of this difference, 
it seems that in the objects with
weak continua, line emission and/or very strongly-rising continua
beyond 13 $\mu$m must provide the bulk of the emission, if the 12
$\mu$m {\em IRAS} fluxes are accurate in this crowded region.

\section{Bulge membership and gas phase abundances}\label{sec:data}

Due to the lack of accurate distance indicators for PNe, bulge
membership cannot be determined with certainty, but it is strengthened
if the following criteria are met:
\begin{itemize}
\item Criterion A. The bulge is the site of large velocity
dispersions, in contrast to the near-circular orbits in the Galactic
disk. If a PN is associated with the Galactic disk, the maximum line
of sight velocity towards the bulge would be of the order of the
velocity dispersion for 1\,M$_{\odot}$ stars in the solar
neighbourhood, or about $\sigma \sim 80$~km\,s$^{-1}$ (from the
age-velocity dispersion relation in Wielen 1977).  In the extreme case
of placing the PNe on the inner side of the molecular ring
(galactocentric radius $R \sim 2.5$~kpc), bulge membership is
strengthened if the deviation from circular rotation is greater than
$\sigma$.
\item Criterion B. PNe with distances beyond 6~kpc are bulge
members. When available, we used the statistical distances from Zhang
(1995), which are averaged between an ionised mass-radius distance
scale, and that in van de Steene \& Zijlstra (1995), based on a
brightness temperature $T_b(6~\mathrm{cm}) -$radius relationship.
Otherwise we used the distances from van de Steene and Zijlstra
(1995), or the distances computed by Maciel (1984, based on the
mass-radius relation). In the case of H1-53, only the Acker et
al. (1991) criterion for bulge membership is available (although H1-53
has an undefined angular size and the flux reported is at 2\,cm).
\end{itemize}
The $D_{IRAS}$ distances discussed in paper~I, based on a constant PN
luminosity of 8500~L$_\odot$ and using the 4 {\em IRAS} bands to
approximate the bolometric flux, cannot be used to identify 
bulge members because of poor quality $IRAS$ fluxes.  However, the
assumption that  50\% of the bolometric flux falls in the 25$\mu$m band
(0.52$\pm$0.1 in the sample of paper~I) yields estimated $D_{IRAS}$
distances that correlate well with the values in
Table~\ref{table:dusttypes_bulge}. The ratio of $D_{IRAS}$ to
6~cm-based distances is 1.9 with a 1$\sigma$ spread of 0.6. The same
ratio is 1.8$\pm$0.8 in the sample of Galactic disk PNe with warm dust
emission. It is interesting to note that, for PNe which are optically
thick to their nuclei's radiation, the approximation of constant
luminosity gives distances which are in excess of 6~cm-based distances
{\em by the same factor} for bulge and disk objects, thereby hinting
at similar PN bolometric luminosity functions (LFs) for the two
populations. Perhaps this is related to the constancy of the 
[O\,{\sc iii}] PNLF in broadly varying galactic enviroments (Ciardullo
et al.  1991)?

Of the C-rich objects, only M\,1-38 satisfies criteria A and B for
bulge membership, and its dust signature is rather uncertain.  M\,3-38
shows a superposition of silicate and UIR emission, and it is
classified as an `O' type PN with the definitions of Table~3 in
paper~I.  We stress that this convention follows the gas-phase
abundances, `O' PNe are also O-rich, although there are only two
objects with such a superposition and known C/O ratio.  Ratag et
al. (1997) report a high N/O ratio for M\,3-38.

Table~\ref{table:dusttypes_bulge} also contains a summary of the
observational data on bulge PNe with 8--13$\mu$m spectra. Nitrogen
enrichment is classified according to Peimbert (1978), with the
sub-types from Faundez-Abans and Maciel (1987) but without the
kinematical distinctions (i.e.  $\log($N/O$) > -0.3$ in type~I PNe,
$-0.3 > \log($N/O$) > -0.6$ in type~IIa PNe, and $-0.6 > \log($N/O$)$
in type~IIb PNe), and the gas-phase abundances are mostly taken from
Ratag et al. (1997).  Note however that the agreement between
different data sets for the few objects in common (Ratag et al 1997,
Cuisinier et al 2000, Webster 1988), which are generally faint and
reddened, is rather poor.  There appears to be a high proportion of
type I PNe in the bulge sample, 25\% of the total, or 30\% of the PNe
which satisfy criteria A and B for bulge membership.  This value is
close to the proportion of type I PNe in compact and infrared bright
Galactic disk PNe,  $29\pm6$\%.  Cuisinier et al. (2000) obtain
similar distributions for N/O ratios in bulge and local PNe,
considering the small number statistics. In fact, although their
interpretation points towards a marked difference with local PNe,
their data shows 23\% type~I objects, and with a continuous range of
N/O values, like the Galactic disk PNe.  There is also a large
population of type IIa nebulae in the bulge, 33\% in this sample,
compared to $\sim$12\% in the disk (in the sample of paper~I).

As an example of the uncertainties affecting the gas phase abundances,
consider the case of M~2-29: ground based spectra yielded
[O/H]$=-1.4$~dex, making M~2-29 the most metal poor PN so far, but
Torres-Peimbert et al. (1997) reported [O/H]$= -0.5$~dex from HST FOS
spectra of a knot close to the central star. Are the HST results
typical of the whole nebula? Would density and temperature
inhomogeneities also affect the abundance determinations of other
bulge PNe?  Although we will consistently refer to the work by Ratag
et al. (1999), the gas-phase abundances can only be viewed as rough
indicators.

The distribution of oxygen abundances listed in
Table~\ref{table:dusttypes_bulge}, for the 10 objects with warm dust
emission and known [O/H], is [O/H]$=-0.3\pm0.4$~dex\footnote{Note we
average logarithmic abundances, so as to avoid negative values when
characterising the distribution of linear abundance values. The
drawback is an `inflating' effect, in this case taking the log of the
linear average of O abundances would give [O/H]$=-0.15$. In this
respect we follow McWilliam \& Rich (1994), and the same procedure is
used in papers~I and II.} (in terms of the mean and 1$\sigma$ spread
and for a solar abundance of $7.4~10^{-4}$, Grevesse \& Anders
1989). In the Galactic disk PNe discussed in paper~I, the distribution
of oxygen abundances for the objects with warm dust emission, is
[O/H]$=-0.52\pm0.26$~dex for $R>R_\circ$, [O/H]$=-0.22\pm0.21$~dex for
$R<R_\circ$, and [O/H]$=-0.31\pm0.26$~dex for all $R$. Thus the mean
[O/H] in bulge and local PNe is very similar, and even if the spread
in [O/H] is larger for bulge PNe, the maximum and minimum observed
values are comparable to local PNe: From [O/H]$\sim -1$~dex to
[O/H]$\sim +0.2$~dex. The distribution of [O/H] values for the PNe
discussed here is very similar to the distribution of [Fe/H] in
Baade's window K giants, $\sim$1\,dex about $<$[Fe/H]$>=-0.3$
(McWilliam \& Rich 1994).

\section{Discussion on the  progenitor masses of PNe towards the bulge}\label{sec:bulge_mod}

The much smaller fraction of C-rich PNe in the bulge compared to the
disk must reflect the change in the stellar populations between the
two regions. The main factors that affect the production of C-rich PNe
are the initial composition and the age of the stellar population; in
old populations the stars sufficiently massive to produce C-rich PNe (
$\ge 1.5$ M$_\odot$) will have evolved beyond the PN stage. In this
Section we examine the consequences of linking the progenitor
population of PNe towards the bulge to the inner Galactic disk or to
the bulge stellar population.

\subsection{Expected frequencies of C-rich PNe for various progenitor populations}\label{sec:bulge_mod_sagb}

It appears that the Galactic metallicity gradient is at the root of
the the trend for an increasing fraction of C-rich PNe beyond the
solar circle (Casassus \& Roche 2001, paper~II), and it could be
thought that a natural explanation for the rarity of C-rich bulge PNe
may be an extrapolation of the Galactic disk trends to even more
metal-rich environments.  However, the distribution of oxygen
abundances listed in Table~\ref{table:dusttypes_bulge} does not reach
the required values.  Also surprising is the fact that C-rich PNe,
common in low metallicity environments, are found at rather high [O/H]
in the bulge sample.

We now examine the consequences of assuming bulge PN progenitors have
diffused off the inner galactic disk, but are otherwise similar to
those of local PNe, and investigate the predictions of single star
evolution using the synthetic AGB model proposed by Groenewegen \& de
Jong (1993), modified to include the results of Forestini \&
Charbonnel (1997, as described in paper~II).

We considered three cases for the initial compositions of the
progenitor population. The first two had initial metallicities
normally distributed about $Z=0.02$ and $Z=0.03$, with an arbitrary
$Z_\mathrm{FWHM}=0.005$.  The third had a distribution following the
observed [O/H] distribution in the objects in our sample i.e. with
$\log(Z/0.02) \sim \mathrm{[O/H]} = -0.3\pm0.4$~dex, and bounded by
$-1$~dex and $+0.2$~dex. Initial compositions were approximated by a
metallicity-scaled solar mix, except for O, for which we also
considered $^{16}\mathrm{O}= ~^{16}\mathrm{O}_{\odot} (Z/0.02)^{1/7}$,
which is obtained by eliminating $R$ from
$\log(Z/0.02)=-0.07(R-R_{\circ})$ and
$\log(^{16}\mathrm{O}/0.02)=-0.01(R-R_{\circ})$ (the oxygen gradient
inferred from B stars, as reported by Smartt et al. 2000).

A power law initial mass spectrum, $p(M_i)dM_i \propto M_i^{-\kappa}
dM_i$, whose exponent $\kappa$ was varied between 2 and 8, is taken to
represent the progenitor population. The mass spectrum of PN
progenitors is given by $N(M) ~dM = \mathrm{SFR} [ t(M) ]~
\mathrm{IMF}(M)~ dM$ where $t(M)$ is the time of birth for a
progenitor of mass $M$.  For a constant star formation rate (SFR), the
Salpeter (1955) initial mass function (IMF) leads to
$\kappa=2.35$. For progenitors linked to the inner Galactic disk, in a
conservative estimate the probability for a given star to diffuse off
the plane would be proportional to the progenitor's lifetime, or about
$M^{-2}$. Assuming randomly distributed star formation bursts behave
on average as a constant SFR gives $\kappa = 4.35$ for a Salpeter
IMF. Actually, diffusion off the galactic disk is probably a
cumulative process, and using the square of the lifetime may be more
adequate, thereby giving $\kappa = 6.35$. Whether there were more/less
bursts in the past than today\footnote{One underlying assumption is
that the diffused population samples all masses, excluding the case of
just one very recent burst (say in the inner 100~pc).  The objects
discussed here are distributed uniformly over the central
10$^{\circ}$.}  can be taken into account by increasing/decreasing
$\kappa$.  If the progenitors are coeval with globular clusters, then
SFR is best approximated as a delta function at $t=0$.  We would have
AGB stars at the $0.8~$M$_{\odot}$ turnoff only (see below), and
$\kappa$ goes to infinity. However, this case is excluded by adopting
a minimum initial mass of 1.2\,M$_{\odot}$ for the progenitors of
compact and IR-bright PNe, as in paper~II, our purpose being to test
whether the bulge and local PNe are similar stellar populations.

\begin{figure}
\begin{center}
 \resizebox{8cm}{!}{\epsfig{file=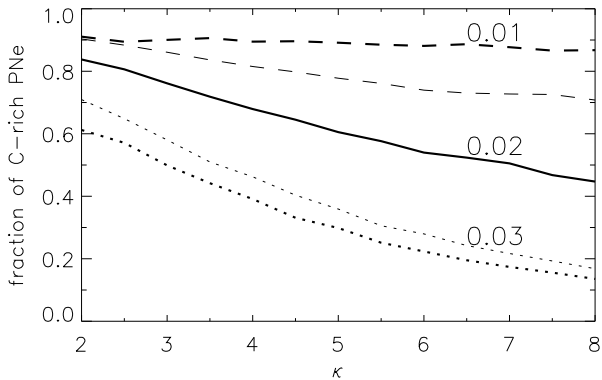}} \caption{The predicted
 fraction of C-rich PNe as a function of $\kappa$, estimated by
 averaging the AGB mass loss composition over the last 2000~yr of
 evolution. The solid line corresponds to $Z=0.02$, the dotted line to
 $Z=0.03$, and the dashed line to the bulge metallicities (a broad
 distribution centred on $Z=0.01$, see text). Thick and thin lines
 correspond respectively to initial compositions as in a $Z$-scaled
 solar mix, or taking into account C/O$(Z)$.}\label{fig:model_bulge}
\end{center}
\end{figure}

The resulting properties of the synthesised bulge PN population can be
seen in Figure~\ref{fig:model_bulge}. The hypothesis that the bulge
and local PNe are similar stellar populations predicts far too many
C-rich objects in the bulge. Taking a $Z$-scaled solar mix (the thick
lines) or C/O$(Z)$ (the thin lines) does not significantly change the
predictions.  It would seem that even for the steepest $\kappa$, the
bulge metallicities predict a majority of C-rich PNe. The only set of
parameters which can account for the observed scarcity of C-rich
objects within 1-$\sigma$ is $Z=0.03$, $\kappa=3.5-6$, i.e. a
super-metal-rich and fairly young population, derived from a constant
SFR and having diffused off the inner disk.

\begin{figure}
\begin{center}
 \resizebox{8cm}{!}{\epsfig{file=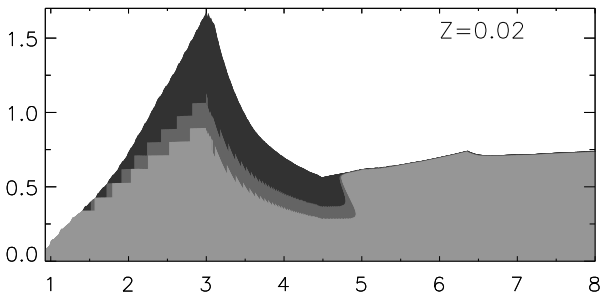}} \caption{ Time spent
on the AGB (10$^{6}$ years, in ordinates) as a function of initial
mass (in M$_{\odot}$), for the model presented in paper~II (at solar
metallicity). The grey scale grows darker from the M to S to C
phases.}\label{fig:tAGB}
\end{center}
\end{figure}

It is also important to calculate the predicted frequency of C stars,
which is observed to be very low in the Galactic bulge (see next
Subsection). The predicted C/M star ratio on the thermally pulsing
AGB, for $\kappa=2$, may be considered an upper limit on the ratio
that would be observed in the synthetic populations: M stars on the
early AGB are bound to contaminate the star counts, and for a steeper
$\kappa$, C stars are increasingly improbable.  For $Z=0.03$, the
predicted ratio is C/M$<0.04$, while for $Z=0.02$ we obtain
C/M$<0.22$, and C/M$<0.35$ for the bulge metallicities. Higher
metallicities are associated with very low values of the C/M star
ratio. The rarity of bulge C stars is naturally explained by their
very short lifetimes, as illustrated in Figure~\ref{fig:tAGB}.  The
very occurrence of this phase is sensitive to the progenitor's mass,
resulting from an interplay of mass loss in the stellar envelope and
the thermal pulses (especially around $\sim 2$~M$_{\odot}$). 


N-enrichment is firmly established to be a function of progenitor
mass, in the case of the Galactic disk population. Thus the statistics
of N/O ratios suggests the bulge PNe are derived from a population
similar in age to local PNe (if they have similar
progenitors). Although the highest N/O ratios, and hence the most
massive progenitors, may be deficient in bulge PNe (Cuisinier et
al. 2000), the upper end of the 1--8\,M$_{\odot}$ mass range is very
infrequent, and derived differences with local PNe should be
negligible (we avoid discussion of the predicted distribution of PN N/O
ratios, as the frequency of N-enrichment in disk PNe is not
understood, see Section~2.2 in paper~II).

For single star evolution to account for the compositions of bulge
PNe, a super-metal-rich and young progenitor population is required,
i.e. an environment typical of the inner Galactic disk. But the oxygen
abundances reported for bulge PNe, taken as one stellar population,
are on average slightly subsolar, and similar to the disk objects.  
Thus, assuming the bulge and local
PNe have similar progenitors (and assuming the reported [O/H] values
are accurate), cannot reproduce the properties of the bulge sample
when taken as one stellar population.

\subsection{Are the bulge PNe all type IV?}\label{sec:discussion_bulge}

The overall picture gathered to date points at an old bulge, composed
of stars no more massive than about 0.8 M$_{\odot}$ with significant
scatter in metallicity about the solar value (Ortolani et al. 1995,
Idiart et al. 1996, Bruzual et al. 1997, Barbuy, Bica \& Ortolani 1998,
Feltzing \& Gilmore 2000). But this result is not unchallenged.
Conclusions preferring a younger bulge have been reached (Holtzman et 
al. 1993, Kiraga, Paczynski \& Stanek. 1997).  In fact, although the
results by Ortolani et al. (1995) have found strong support, they 
refer to the bulk of the bulge; some observations suggest the
existence of an intermediate mass population\footnote{ Mc\,William and
Rich (1994) estimate that the average mass of K giants in Baade's
window can be no less than 1.1 M$_{\odot}$. The distribution of 
periods of LPV stars in the bulge has been interpreted as evidence 
for a population with ZAMS masses in
excess of 1.3~M$_{\odot}$ (Harmon and Gilmore 1988), but this has
recently been shown by Frogel \& Whitelock (1998) to merely reflect a
metallicity effect in an old population.}. Frogel (1999) summarises his
work on a bulge population less than $\sim$1\,Gyr old: although there
is no solid evidence for a young population in Baade's window, a mix
of young and old populations is found towards the inner bulge (fields
less than 1$^{\circ}$ from the Galactic centre), as evidenced by an
increase in luminous AGB stars.

A long standing problem related to the existence of an intermediate
mass population in the bulge is the lack of C stars. Blanco, Blanco \&
McCarthy (1978) detected only one C star in 310 M stars.  Azzopardi, 
Lequeux \& Rebeirot (1991) reported a list of 33 C stars detected
towards the bulge, but their properties are puzzling. Although the
C stars share the same kinematics as bulge K and M giants, they are
too faint to be on the AGB.  The bulge C stars do not seem to be
genuine AGB stars, and could be dwarf C stars (dCs) along the line of
sight to the bulge. dCs are observed to have low metallicities 
(Harris et al. 1998), and are most
likely produced through binary evolution with $Z<0.5$\,Z$_{\odot}$
(De\,Kool \& Green 1995), making them members of the local population
of spheroid dwarfs. Thus the lack of genuine AGB C stars is
characteristic of the bulge.

The principal piece of information brought by the warm dust emission
features of PNe in the direction of the bulge is that the proportion
of C-rich nebulae is dramatically lower, only $\sim$30\% against
$\sim$78\% in local PNe. In an extension of the method applied to disk
PNe in papers I and II, we showed in section \ref{sec:bulge_mod_sagb}
that such a low fraction of C-rich nebulae and the nitrogen enrichment
reported by Ratag et al. (1997) require a young and metal-rich
population, with Z$\sim$0.03. Yet on average the bulge and local PNe
have similar metallicities, although the four C-rich object in the
bulge sample are found at rather high [O/H], and none is a definite
bulge member.

Two alternatives should be considered: 
\begin{enumerate}
\item The bulge PNe with warm dust could be related to the type IV PNe
in Peimbert's classification, i.e. halo PNe with peculiar
abundances. The case of BB-1 (or PN G108.4-76.1), studied by Pe\~{n}a
et al. (1993), merits attention as it is metal poor (O is underabundant
relative to solar  by one order of magnitude), has a velocity of
$\sim$200~km\,s$^{-1}$, and is nonetheless enriched in carbon and
nitrogen, $\log$(C/O)=+1, $\log$(N/O)=+0.2.
\item The bulge PNe with warm dust are similar to the local
population, but with higher metallicity and linked to the inner
regions of the Galactic disk. A mechanism remains to be found by which
such a population would diffuse off the Galactic plane. 
\end{enumerate}

Aspects of both alternatives seem applicable to the bulge PNe. The
C-rich PNe are most likely metal-rich and foreground objects, and the
lack of counterpart C stars is natural in metal-rich environments (the
C phase would occur only at the very end of AGB evolution, and could
possibly be linked with PN ejection). 

Another piece of information is provided by the lack of metallicity
dispersion in Baade's window M giants, about a mean slightly larger
than the solar value (Frogel \& Whitford 1987, Terndrup et
al. 1991). The dispersion in metallicity for M giants is at most a
factor of 2 in [Fe/H], about $<$[Fe/H]$>=+0.3$, in contrast with the
large spread in metallicity of Baade's window K giants, $\sim$1\,dex
in [Fe/H] about $<$[Fe/H]$>=-0.3$.  Is the bulk of the M giants
observed towards Baade's window from a young population, with uniformly
high metallicity, linked to the inner Galactic disk?

Similar to the superposition of the populations traced by the K and M
giants in Baade's window, there could be a superposition of two
population in the bulge PNe, one linked to the inner disk and the
other to the old bulge.

\section{Conclusion}\label{sec:conc_bulge}

The C/O chemical balance for a sample of PNe towards the Galactic bulge 
has been established through the 8--13\,$\mu$m dust emission features. Out
of 15 PNe with identified dust types, 4 PNe have C-based grains. The very
low fraction of C-rich PNe is in contrast with the local PNe, and the
bulge PNe probably form a different population. However, none of the
C-rich PNe is an unquestionable bulge member. The fraction of C-rich
nebulae among `bona fide' bulge PNe could be much lower.

In the framework of single-star evolution, the properties of the bulge
PN sample appear to be derived from a young and metal-rich population,
linking the `bulge' PNe to the inner Galactic disk. However, the
distribution in metallicity of the bulge warm dust PNe follows that of
Baade's window K giants, with a similar mean to solar neighbourhood PNe.
It is difficult to reconcile the C/O balance inferred from the dust
emission with a single bulge PN population. 
We are thus inclined to follow the suggestion by Peimbert (1992) that the
bulge PNe are predominantly type~IV, although contamination from
Galactic disk objects is likely.  In this context the C-rich
PNe towards the bulge are linked to the inner Galactic disk, and the lack
of counterpart C stars is a natural property of metal-rich
environments. 

We find that M\,2-29, usually classified as a type IV PN with
Z$\sim$0.0008, shows silicate emission. It seems, however, that the
ground-based [O/H] determinations could be in error (Torres-Peimbert
et al. 1997). In the light of the significant uncertainties linked
with the gas-phase abundances, discriminating between a bulge or a
disk origin for `bulge' PNe based on their [O/H] values may be
prematurate. It would be interesting to obtain 8--13\,$\mu$m spectra
of the 7 type IV PNe listed by Peimbert (1992) to investigate dust
production at low metallicities.


\section*{Acknowledgments}

We are grateful to the referee for an interesting
report. S.C. acknowledges support from Fundaci\'{o}n Andes and PPARC
through a Gemini studentship.

\bsp
\label{lastpage}
\end{document}


\small
\setcounter{table}{1}
\begin{center}
\begin{sidewaystable}
\caption{List of PNe in the direction of the galactic bulge with
8-13$\mu$m spectra, and the identified warm dust emission features.} 
\label{table:pnlist_bulge}

\begin{tabular}{lrclcccrrclll} \\
       &   PN~G	        & dust& N/O    &He$^{a}$    &N$^{a}$ &O$^{a}$ &$V_\mathrm{lsr}$  & $d$ &  bulge &ref$^{c}$ & comments$^h$    \\
       &                & type&	type   &      &      &     &         km\,s$^{-1}$        &  kpc   & PN?$^{b}$ &       &       \\ \hline
       &       	        &     &	       &      &      &     &     	                 &          &           &  & \\
H1-54      & $  2.1-4.2$&  O  & IIb &10.99 & 7.48 & 8.56&$-116  $&11.4&   AB &4;2 & silicates:1,191                  \\
M2-23      & $  2.2-2.8$&  O  & IIb &10.98 & 7.40 & 8.22&$ 186  $&8.4 &   AB &1;2 & silicates $^{e}$                  \\
H1-63      & $  2.3-6.4$&  O  & x   &-     & -    & -   &$-1.0  $&$<$5.7&    &4;6 & silicates:1,195           \\
Cn1-5      & $  2.3-9.5$&  C  & IIa &11.06 & 8.63 & 9.05&$-29.7 $&4.4 &      &4;2 & graphite:0.15,130 UIR:0.58,0.15,0.12 \\
M1-38      & $  2.5-3.7$&  C  & IIb &11.04 & 8.21 & 9.00&$-70.0 $&6.6 &   AB &4;2 & graphite:0.4,144 UIR:---,0.3,0.3   \\
M1-37      & $  2.6-3.4$&  +  & I   &11.10 & 8.10 & 7.98&$ 220  $&8.1 &   AB &1;1 & weak continuum$^{e}$              \\
M1-35      & $  3.9-2.3$&  +  & I   &11.35 & 8.79 & 8.48&$ 67.4 $&4.6 &      &1;2 & weak continuum   \\
M2-29$^{d}$& $  4.1-3.0$&  O  & IIa &11.11 & 7.01 & 7.47&$-112  $&9.9 &   AB &1;2 & silicates:1,215           \\
H1-53      & $  4.3-2.6$&  O  & x   &-     & -    & -   &$ 75.0 $&-   &      &4;5 & silicates:1,199           \\
M1-25      & $  4.9+4.9$&  +  & I   &11.10 & 8.82 & 9.09&$ 14.1 $&4.9 &      &1;2 & weak continuum            \\
M1-20      & $  6.2+8.4$&  c  & IIb &11.02 & 7.75 & 8.62&$ 60.3 $&5.9 &      &1;2 & graphite:0.6,191 SiC:0.4,120              \\
M3-15      & $  6.8+4.1$&  +  & IIa &11.03 & 8.14 & 8.74&$ 97.5 $&4.1 &      &1;2 & weak continuum            \\
NGC6644    & $  8.3-7.3$&  +  & IIa &11.03 & 8.00 & 8.54&$ 194  $&4.1 &   A  &4;2 & weak continuum $^{e}$             \\
M3-21      & $355.1-6.9$&  +  & IIa &11.05 & 8.26 & 8.74&$-66.1 $&5.6 &      &1;2 & weak continuum                    \\
H1-32      & $355.7-2.7$&  O  & IIa &10.95 & 7.59 & 8.12&$-222  $&-   &   A  &3;- & silicates:1,228                   \\
M1-27      & $356.5-2.4$&  C  & IIb &-     & 8.22 & 8.86&$-49.7 $&3.7 &      &1;2 & graphite:0.6,220 SiC:0.4,194$^{f}$                \\
M2-24      & $356.9-5.8$&  +  & x   &-     & -    & -   &$145.5 $&-   &   A  &7;- & unidentified $^{g}$\\
M3-38      & $357.0+4.4$&  O  & I   &11.11 & 8.69 & 8.37&$-156  $&8.3 &   AB &4;2 & graphite:0.21,296 silicates:0.52,152 UIR:---,---,0.27  \\
M3-8       & $358.2+4.3$&  O  & I   &11.11 & 8.75 & 9.02&$ 95.0 $&7.5 &   AB &4;2 & silicates:1,291   \\
M4-6       & $358.6+1.8$&  +  & IIb &10.98 & 7.97 & 8.31&$-292  $&6.3 &   AB &4;2 & weak continuum$^{e}$              \\
Th3-26     & $358.8+3.0$&  +  & x   &-     & -    & -   &$ 204  $&7.9 &   AB &4;2 & weak continuum    \\
M1-29      & $359.1-1.7$&  +  & I   &11.14 & 8.82 & 9.00&$-41.3 $&3.3 &      &1;2 & weak continuum            \\
M3-44      & $359.3-1.8$&  +  & x   &-     & -    & -   &$-89.0 $&$>$1&   A  &4;6 & weak continuum $^{e}$             \\
H1-40      & $359.7-2.7$&  O  & IIa &11.06 & 8.08 & 8.70&$ 63.8 $&7.8 &   AB &1;1 & silicates $^{e}$                  \\
M2-27      & $359.9-4.5$&  +  & IIa &11.07 & 8.46 & 8.77&$ 145 $&5.7 &   A  &1;2 & weak continuum$^{e}$              \\                    \hline
\end{tabular}

\medskip
$^{a}$ The abundances were taken from Ratag et al. (1997), except for
M1-27 which is from Cuisinier et al. 2000. 

$^{b}$ Criteria met for bulge membership (see text). 

$^{c}$ References for measurements of V$_\mathrm{lsr}$ and estimates  
of the distance respectively. The codes are:  1)
Zijlstra, Acker \& Walsh (1997); 2) Zhang (1995); 3) Kohoutek and Pauls
(1995); 4) Schneider et al. (1983); 5) Acker et al. (1991); 6) Maciel
(1984); 7) Durand, Acker \& Zijlstra  (1998).

$^{d}$ The abundances for M~2-29 are uncertain: The analysis by
Torres-Peimbert et al. (1997) suggests previously reported [O/H]
values are underestimated. 

$^e$ 8--13 $\mu$m spectra from Roche  (1987).

$^f$ The fit to M\,1-27 with SiC and graphite had $\chi^{2}$=1.03, while graphite alone gave $\chi^{2}$=1.9, and a trapezium emissivity function resulted in
$\chi^{2}$=2.5.

$^g$ M\,2-24's 8--13~$\mu$m spectrum is atypical of PNe (compare with
K\,4-57 in paper~I).

$^{h}$ Best fit parameters for the 8--13$\mu$m continua:
(graintype):$a^{\prime},T$ for SiC or silicates, where $a^{\prime}$ is the
relative contribution and $T$ the temperature of the modified black
body corresponding to (graintype); UIR:a$^{\prime}_{7.7}$,
a$^{\prime}_{8.6}$, a$^{\prime}_{11.3}$ for the unidentified IR bands (see
Section~2 for more details).

\end{sidewaystable}
\end{center}